\documentclass[conference]{IEEEtran}
\IEEEoverridecommandlockouts
\usepackage{cite}
\usepackage{amsmath,amssymb,amsfonts}
\usepackage{algorithmic}
\usepackage{graphicx}
\usepackage{textcomp}
\usepackage{enumitem}
\usepackage{xcolor}
\usepackage{color}
\usepackage{subcaption}
\def\BibTeX{{\rm B\kern-.05em{\sc i\kern-.025em b}\kern-.08em
    T\kern-.1667em\lower.7ex\hbox{E}\kern-.125emX}}
\begin{document}

\title{Performance Modeling and Analysis of a Hyperledger-based System Using GSPN\\}

\author{\IEEEauthorblockN{Pu Yuan\IEEEauthorrefmark{1},
		Kan Zheng\IEEEauthorrefmark{1},
		Xiong Xiong\IEEEauthorrefmark{1}, Kuan Zhang\IEEEauthorrefmark{2} and
	    Lei Lei\IEEEauthorrefmark{3}}
    \IEEEauthorblockA{\IEEEauthorrefmark{1}
    	Intelligent Computing and Communications Lab, Key Lab of Universal Wireless Communications\\
    	Beijing University of Posts and
    	Telecommunications\\
    	Beijing 100876, China\\}
    \IEEEauthorblockA{\IEEEauthorrefmark{2}
    	Department of Electrical and Computer Engineering \\
        University of Nebraska–Lincoln\\
    	Omaha, NE 68182, USA.\\}
   \IEEEauthorblockA{\IEEEauthorrefmark{3}
    	College of Science and Engineering\\
 	    James Cook University\\
     	Douglas, QLD 4814, Australia\\
		Email: yp2013@bupt.edu.cn,
	           ajodfaj@bupt.edu.cn,
	           kuan.zhang@unl.edu,
		       lei.lei@jcu.edu.au\\
		Corresponding author: zkan@bupt.edu.cn}}

\maketitle

\begin{abstract}
As a highly scalable permissioned blockchain platform, Hyperledger Fabric supports a wide range of industry use cases ranging from governance to finance. In this paper, we propose a model to analyze the performance of a Hyperledger-based system by using Generalised Stochastic Petri Nets (GSPN). This model decomposes a transaction flow into multiple phases and provides a simulation-based approach to obtain the system latency and throughput with a specific arrival rate. Based on this model, we analyze the impact of different configurations of ordering service on system performance to find out the bottleneck. Moreover, a mathematical configuration selection approach is proposed to determine the best configuration which can maximize the system throughput. Finally, extensive experiments are performed on a running system to validate the proposed model and approaches.
\end{abstract}

\begin{IEEEkeywords}
blockchain, Hyperledger Fabric, performance modeling, Generalised Stochasitc Petri Nets
\end{IEEEkeywords}

\section{Introduction}
Blockchain technology originated from Bitcoin has been growing rapidly in recent years \cite{b1}. The blockchain leverages cryptographic techniques, distributed ledgers and consensus algorithms to provide a trusted and decentralized service for several applications \cite{b2,b3,b4,b5,b6}. Depending on the user authorization mechanisms, blockchain can be mainly categorized  into the permissionless and the permissioned blockchain \cite{b7}. Ethereum is a programmable permissionless blockchain platform that achieve business logic based on specific smart contracts \cite{b8,b9}. On  the other hand, Hyperledger Fabric is an open source enterprise-grade permissioned blockchain platform with a highly modular and configurable architecture \cite{b10}. It integrates fine-grained access control, immutable ledger and pluggable consensus protocols. Due to those advantages, Hyperledger Fabric is used in many scenarios, such as emission trading, insurance and education \cite{b11,b12,b13,b14}.

The performance of Hyperledger Fabric was widely studied \cite{b15,b16}. The performance differences between Hyperledger Fabric version 0.6 and version 1.0 have been evaluated in \cite{b17}, which indicated that version 1.0 can significantly improve the system performance. A repeatable evaluating methodology was proposed to assess the performance of Hyperledger Fabric and Ethereum in \cite{b18}. The consensus protocols of those two platforms were compared in \cite{b19}, where Hyperledger Fabric achieves the better performance than Ethereum. Moreover, extensive experiments with varying parameters on Hyperledger Fabric v1.0 have been conducted to study the impact of various system configurations \cite{b20}. The experimental results indicated that the endorsement policy verification, sequential policy validation of transactions in a block and the state validation and commit are three major performance bottlenecks. Based on those analyses, several optimizations to improve the overall performance were introduced.

However, those experiment-based analyses for Hyperledger Fabric are lack of scalability and theoretical basis. In order to further analyze the performance characteristics of this blockchain framework, it is imperative to model the transaction flow of it using a mathematical approach. Due to the complexity of the transaction flow and system configurations, the performance of Hyperledger-based system is affected by several factors. Thus, there are many difficulties in modeling the system.

In this paper, we analyze the blockchain performance based on Hyperledger Fabric. As a formal mathematical theory designed for modeling concurrency, causality and conflict, GSPN provides a graphical approach to decompose the request processing flow of the Hyperledger-based system into multiple phases. Moreover, the GSPN-based model can be simulated to obtain the performance metrics such as latency and throughput of each phase at a non-steady state, which is suitable for this focused scenario.To sum up, our major contributions are summarized as follows: i.e.,
\begin{enumerate}
\item An analytical model is proposed to depict the transaction flow of a Hyperledger-based system using GSPN and validate this model by experiments.
\item We analyze how different ordering strategies affect the system performance and identify the performance bottleneck based on the proposed model.
\item In response to the identified bottleneck above, a mathematical configuration selection approach is proposed to determine the configuration parameters of the ordering service in order to achieve the maximum throughput of this system.
\end{enumerate}
To validate the proposed model and configuration selection approach, a running system is setup on a cloud server. Furthermore, our work has important guiding significance for the practical use of the Hyperledger-based systems.

The remainder of this paper is organized as follows. In Section II, we investigate the related work. Section III introduces a Hyperledger-based system and the transaction flow of it. Then an analytic model based on GSPN and a configuration selection approach to achieve best system performance are proposed in Section IV. Next, Section V validates our model and approach by conducting extensive experiments on a running system. Finally, Section VI outlines the main conclusions.

\section{Related Work}
Compared with those experiment-based performance analyses mentioned before, the performance modeling for Hyperledger Fabric is more critical and scalable to analyze the characteristics of this blockchain framework. To depict the system accurately, an appropriate mathematical theory is needed. As a modeling approach, Petri Nets is a formal mathematical theory with rigorous mathematical foundation and intuitive graphical representation. Derived from Petri Nets, the Stochastic Petri Nets (SPN) associates an exponentially distributed delay with the firing of each transition to provide a clear and intuitive formalism for generating Markov processes. Based on SPN, the Generalised Stochastic Petri Nets (GSPN) adds immediate transitions and inhibitor arcs to prevent the model from becoming exceedingly large. Moreover, the Stochastic Reward Nets (SRN) introduces more primitives than GSPN to enhance the expressive ability. All those nets are widely used in system performance modeling and analysis \cite{b21,b22,b23,b24}. As for modeling for Hyperledger Fabric, a SRN model of the PBFT consensus based on Hyperledger Fabric v0.6 was presented to discuss how the number of peers affects the consensus latency in \cite{b25}. Furthermore, an overall performance model of Hyperledger Fabric v1.0+ was proposed in \cite{b26}, which discussed the impact of different parameters on system performance such as latency, throughput and utilization. The proposed model was validated by using a test framework named Hyperledger Caliper. However, these studies only analyzed the performance characteristics based on simulation but they did not provide a method to obtain the appropriate configuration parameters of system.

\section{System Overview}
\subsection{System Architecture}
A Hyperledger-based emission trading system is proposed to solve the defects of existing centralized systems \cite{b9}. The system uses two organizations to represent the environmental agency and the trading center respectively and share the same ledger through one single channel. By integrating the characteristics of blockchain, the polluters can achieve the credible trading service through this system. As a typical implementation of Hyperledger Fabric v1.2, the system architecture of the blockchain system is given in Fig. 1.
\begin{figure}[htbp]
	\centerline{\includegraphics[width=\linewidth]{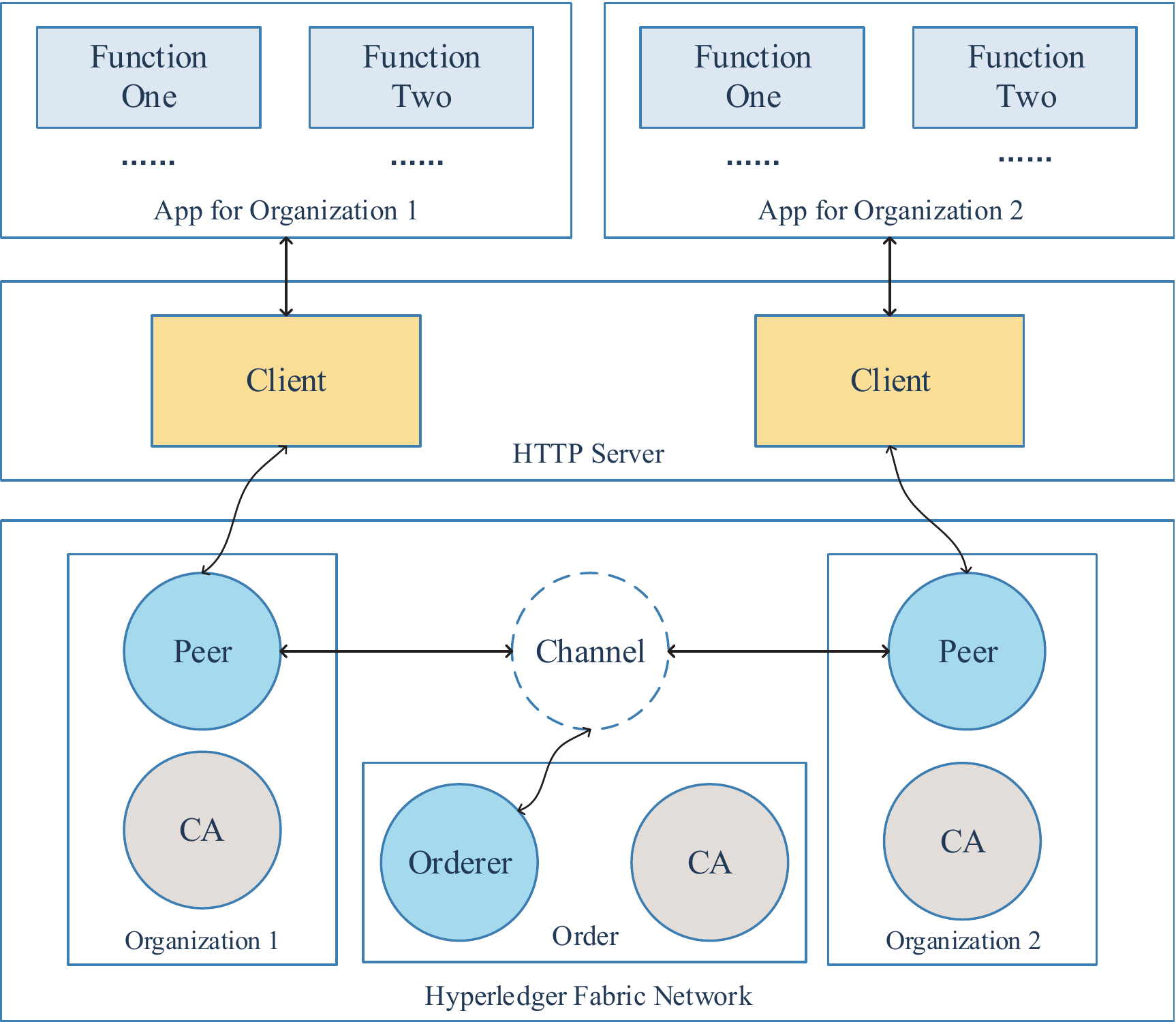}}
	\caption{System Architecture.}
	\label{fig1}
\end{figure}

This system consists of HTTP server, Web application and the blockchain network. The HTTP server plays a role of Fabric client to interact with the underlying network by integrating specific Fabric SDK. Depending on those RESTful APIs provided by the HTTP server, the web application can provide a variety of services for users. The blockchain network contains two distinct organizations, one channel which connects those organizations and an orderer node to  provide the ordering service. This node adopts solo consensus protocol to guarantee the consistency of the distributed ledger. Each organization has a Fabric CA and a local peer. Fabric CA issues certificates for participants to achieve the access control policy. The peer holds an immutable ledger based on LevelDB and installs customized chaincode which implements the business logic. Each peer in this system plays roles of not only an endorser (endorse for transactions) but also a committer (hold ledgers). In addition, all of those nodes are running in special docker containers.

\subsection{Transaction Flow}
Requests in Hyperledger Fabric are divided into kinds, i.e., query and invoke,  depending on whether the ledger is modified. Fig. 2 depicts the transaction flow of a typical invoke-type request. A completed transaction consists of several phases as follows, i.e.,

\begin{figure}[htbp]
	\centerline{\includegraphics[width=\linewidth]{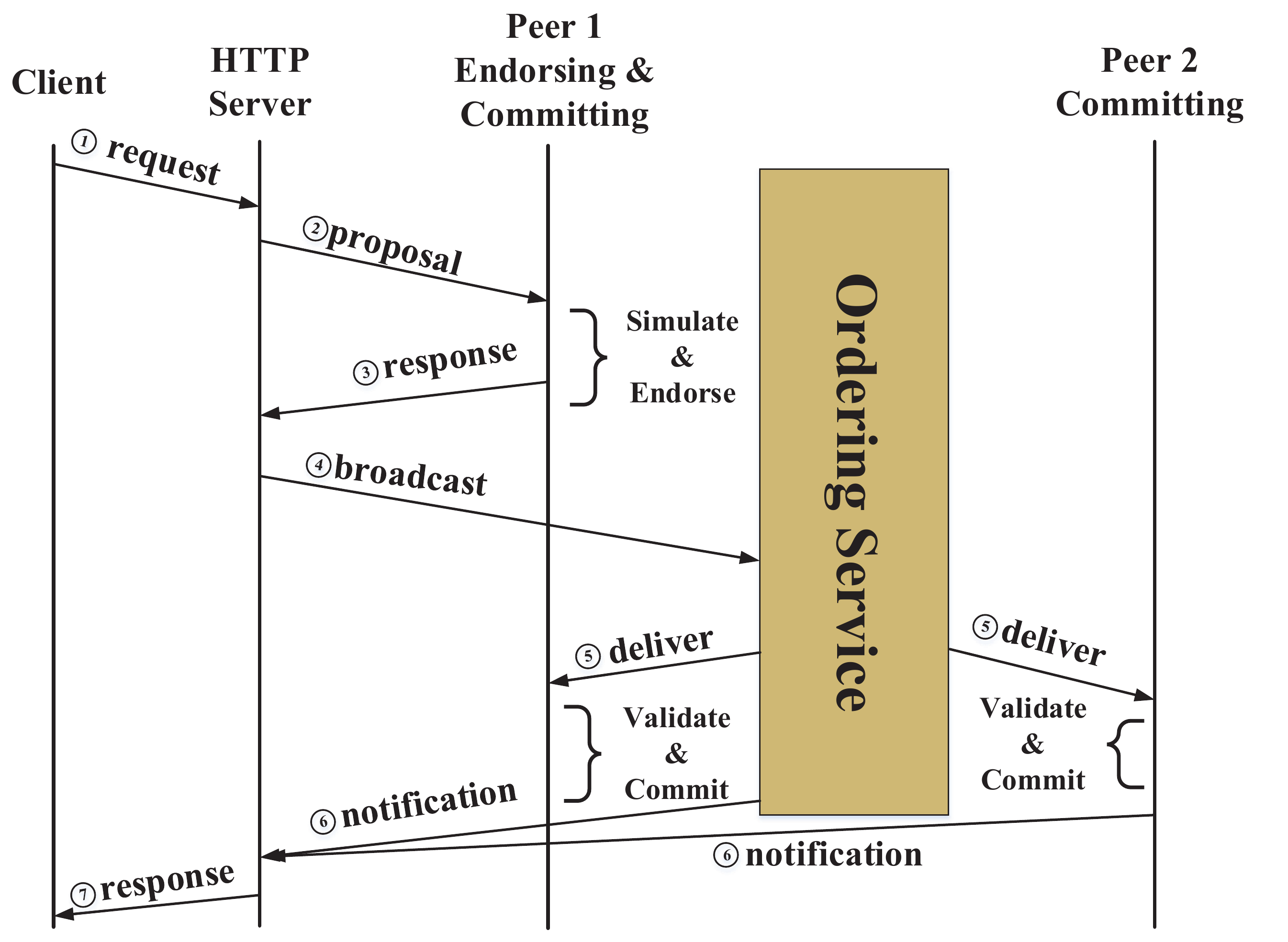}}
	\caption{Transaction Flow.}
	\label{fig2}
\end{figure}

\textbf{HTTP Phase.} Client sends a HTTP request to the server, aiming to interact with the blockchain network. The HTTP server extracts essential parameters from the request body and then constructs a transaction proposal by using the provided SDK. The generated proposal is signed with the client’s credentials and contains details of the specific chaincode. Then, the proposal is sent to an endorsing peer to endorse for this transaction.

\textbf{Endorsement Phase.} All the peers that have installed the chaincode can play the role of an endorsing peer. When receiving a transaction proposal, each endorsing peer should execute tasks as follows: Firstly, the peer verifies the identity of this submitter to check whether it’s authorized to invoke the chaincode. Secondly, the peer executes the chaincode to generate the response value and read-write set without modifying the world state. Thirdly, endorser signs the proposal response with its identity and then the peer sends the response back to the client. Finally, the client collects responses from multiple endorsers and verifies if they satisfy the endorsement policy. The system adopts ‘or’ policy so that one endorser is enough. 

\begin{figure*}[htbp]
	\centerline{\includegraphics[width=\linewidth]{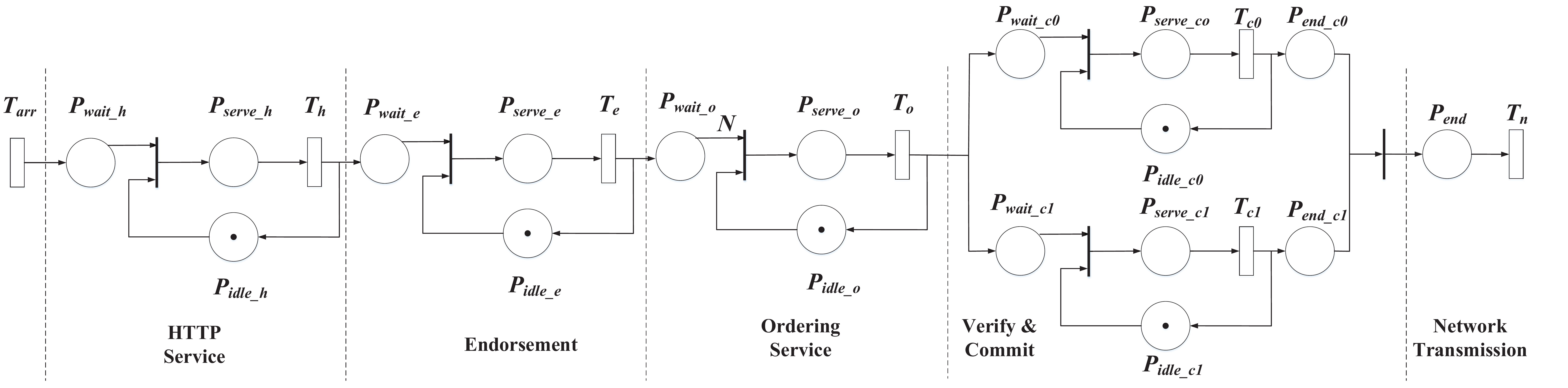}}
	\caption{GSPN Model of System.}
	\label{fig3}
\end{figure*}

\textbf{Ordering Phase.} Once the transaction is fully endorsed, the client integrated in HTTP server broadcasts it to the ordering service. According to the specific configuration, the ordering service orders all the transactions and packages them into blocks with the special strategy. Then it signs the blocks and delivers them to the leading peers using gossip protocol. Some of the critical parameters of ordering service are listed as follows, i.e.,

\begin{itemize}
	\item \textbf{BatchTimeout:} The amount of time to wait before creating a batch.
	\item \textbf{MaxMessageCount:} The maximum number of messages batched into a block.
	\item \textbf{PreferredMaxBytes:} The maximum number of bytes allowed for the serialized messages in a block.
\end{itemize}

When one of the above conditions is satisfied (e.g., the number of transactions queuing in ordering service reaches the MaxMessageCount), a sequence of transactions can be batched into a new block. Considering that the size of transaction is similar, in general only BatchTimeout and MaxMessageCount need to be adjusted.

\textbf{Validation \& Committing Phase.} After receiving blocks from the ordering service, the leading peers disseminate those blocks to all the peers, which belong to the same channel and organization. The peers first verify the signature of the blocks and then check all the transactions within them. If all the transactions pass the endorsement validation and read-write set validation, those blocks are appended to the ledger and the world states are updated.

\textbf{Response Phase.} By registering an event listener in ChannelEventHub, the HTTP server can receive a notification when the target transaction has been committed into a block and appended to the ledger. A registered callback function can collect details of this event and form the response data in JSON format.

\section{The GSPN-based Analytic Model of System}
In this section, an analytic model for the Hyperledger-based system mentioned in Section III is proposed. We first introduce the basic elements of GSPN and the modeling assumptions. Then the proposed model is decomposed to multiple phases and each phase is described in details. Finally, we present a configuration selection approach to determine the network parameters.

\subsection{Preliminaries}
A typical GSPN model consists of basic elements as follows, i.e., 

\textbf{Places.} Circular nodes are used to describe places, which represent conditions or local system states.

\textbf{Transitions.} Rectangular boxes are used to describe transitions, which represent events occur in the system. The fire of a transition can change system status from one place to another.

\textbf{Tokens.} Black dots or numbers are used to describe the tokens resided in place, which represent the state quantity that a place holds. 

\textbf{Arcs.} Arcs specify the relationship between places and transitions. The weight of input arc represents the number of tokens consumed by the transition firing, and the weight of output arc represents the number of tokens produced to the output place. The default weight of arc is 1.

\textbf{Immediate Transitions.} Thick bars are used to describe immediate transitions, which represent events that are assumed to take no time.

\textbf{Inhabit Arcs.} An inhibitor arc from a place to a transition means the transition cannot fire if there is a token in the place.

Based on those basic elements, a performance model for the system mentioned in Section III can be proposed. The system modeling satisfies the following assumptions, i.e., 

\begin{itemize}
	\item The arrival of requests is a Poisson process.
	\item The size of each transaction is constant.
	\item Ignore the transmission latency between processing phases.
\end{itemize}

\subsection{Model Description}
The proposed analytic model based on GSPN is shown in Fig. 3. The overall model depicts the transaction flow in details, consisting of five phases as follows, i.e.,

\textbf{HTTP Phase.} This phase represents the process by which the HTTP server receives the request and sends the transaction proposal to the endorser. It can be regarded as a fundamental processing unit. Fig. 4 decomposes this unit from the overall model. 

\begin{figure}
	\centerline{\includegraphics[width=\linewidth]{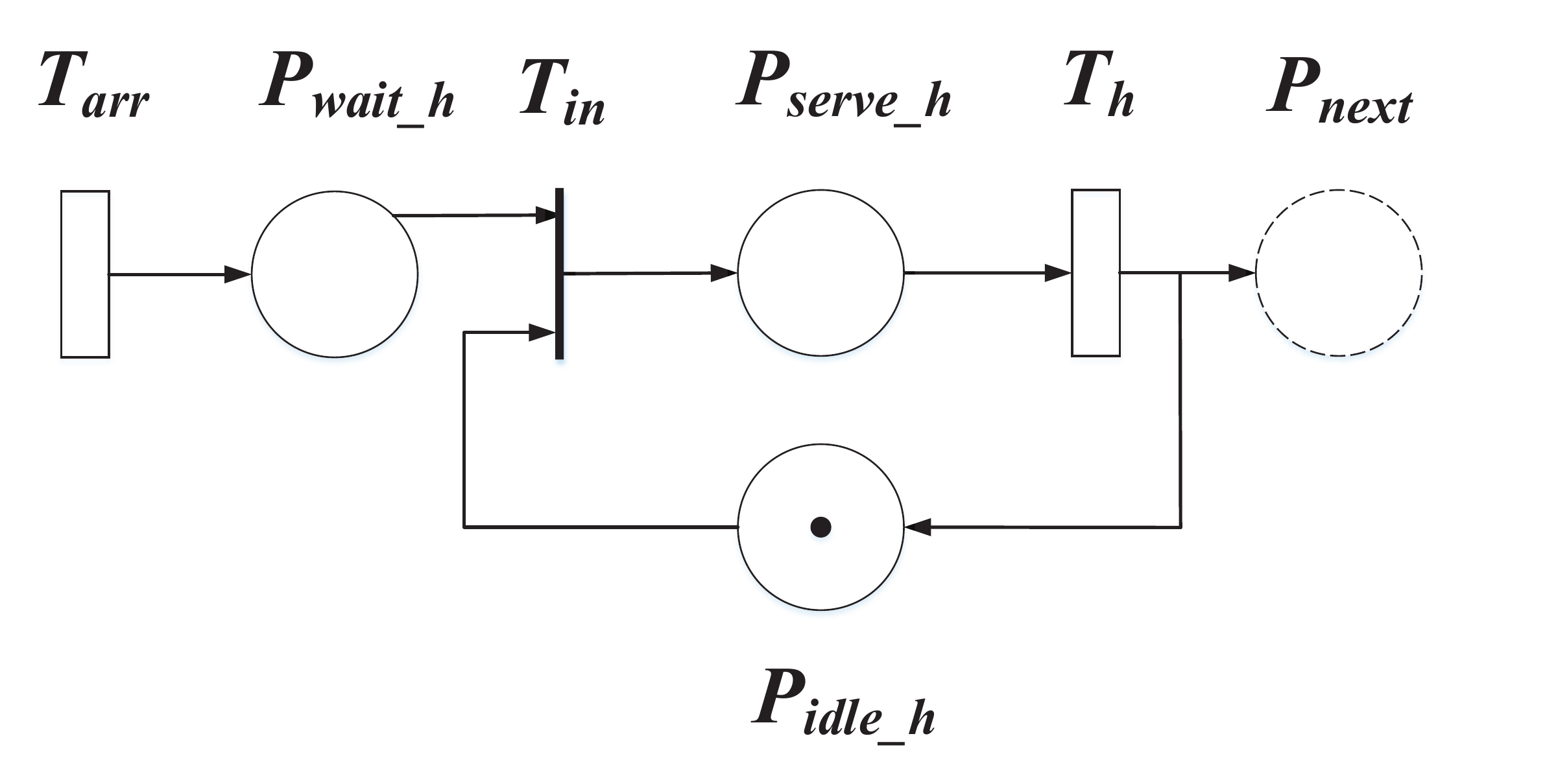}}
	\caption{Decomposed model of HTTP server.}
	\label{fig4}
\end{figure}

The meaning of all the places and transitions are described as follows, i.e.,

\begin{itemize}[leftmargin=1.5cm]
	\item[$T_{arr}$:] a transition that represents the arrival of a new request.
	\item[$P_{wait\_h}$:] a place that represents the request is queuing, the number of token $\#(P_{wait\_h})$ denotes the queuing length.
	\item[$P_{serve\_h}$:] a place that represents the request is being processed.
	\item[$P_{idle\_h}$:] a place that represents the server is idle now, the number of token $\#(P_{idle\_h})$ denotes the number of idle servers.
	\item[$T_{in}$:] an immediate transition whose enable predicate is $\#(P_{wait\_h})>0$ \& $\#(P_{idle\_h})>0$, which means there are idle servers and queuing requests.
	\item[$P_{next}$:] a place represents the next processing phase.
\end{itemize}

The performance metrics of each place in this model can be obtained by simulating. Assume that during the total simulation time $T$, the transition $T_{arr}$ has been fired $X$ times, i.e., place $P_{wait\_h}$ generates $X$ tokens. The firing time interval obeys exponential distribution with parameter $\lambda$. Let $\varphi_p$ denote the number of departed tokens of place $P$, $\tau_{p_i}$ denote the stay time of $i-th$ token in place $P$. The performance metrics of place $P$ can be derived as follows, i.e.,
\begin{itemize}
	\item Throughput: \begin{equation}
	\theta_p = \frac{\varphi_p}{T}.
	\end{equation}
	\item Average Latency: 
	\begin{equation}
	L_P = \frac{\sum_{i}\tau_{p_i}}{\varphi_p}.
	\end{equation}
	\item Queuing Length:
	\begin{equation}
	Q_P = \theta_P * L_P = \frac{\sum_{i}\tau_{p_i}}{T}.
	\end{equation}
\end{itemize}

Obviously, this processing unit describes a typical M/M/1 queuing model, whose total latency equals to the queuing time plus the service time, i.e.,
\begin{equation}
\delta_H = L_{P_{wait\_h}} + L_{P_{serve\_h}}.
\end{equation}

\textbf{Endorsement Phase.} This phase describes the endorsement process in the processing flow. Since the system adopts the ‘or’ endorsement policy, one endorser for a request is enough. So there is only one fundamental processing unit in this phase. According to the above analysis, the total latency of this phase $\delta_E$ can be derived.

\textbf{Ordering Phase.} This phase describes the ordering service of the system. Because the block packaging, signing and delivering are executed sequentially, only one place $P_{serve\_o}$ is used to represent the processing of the orderer node. Different with the above phases, the arc between place $P_{wait\_o}$ and immediate transition $T_{in}$ has a weight $N$, which indicates that the enable predicate of $T_{in}$ is $\#(P_{wait\_o})\ge N$ \&  $\#(P_{idle\_o})>0$. When $T_{in}$ fires , $N$ tokens in $P_{wait\_o}$ are absorbed and one token is created in $P_{serve\_o}$. Corresponding to the actual system, this weight represents the operation of packing N transactions into a block. Similarly, the total latency of this phase $\delta_O$ can be derived.

\textbf{Committing Phase.} When the peer node receives a new block, it performs a series of validations (e.g., MVCC) and then commits the block into the local ledger. Those processes are abstracted into one place $P_{serve\_c}$ to simplify the model. Since two peers in the system perform commit operations synchronously, the HTTP server listens to the events of both peers at the same time. Thus,  there are two parallel procession units in this phase. The total latency of this phase depends on the one with the larger latency.
\begin{equation}
\delta_C = \max(L_{P_{wait\_c0}} + L_{P_{serve\_c0}},L_{P_{wait\_c1}} + L_{P_{serve\_c1}}).
\end{equation}

\textbf{Response Phase.} In order to approximate the real system,  an extra transition is used to represent the network latency between applications and HTTP server,  which can be  expressed as $\delta_T=L_{P_{end}}$.

The total latency of this system can be derived by
\begin{equation}
\Delta=\sum_{i=H}^{T}\delta_i.
\end{equation}

Moreover, it is obvious that a request has been completely handled after it arrives at the place $P_{end}$. Therefore, the throughput of this system is equal to the throughput of place $P_{end}$, i.e.,
\begin{equation}
\Theta =\theta_{P_{end}}.
\end{equation}

Based on the proposed analytic model, the system performance metrics can be obtained conveniently through the simulations after determining the rates of all transitions.
\begin{figure*}[ht]
	\centering
	\begin{subfigure}[b]{0.32\textwidth}
		\includegraphics[width=\textwidth]{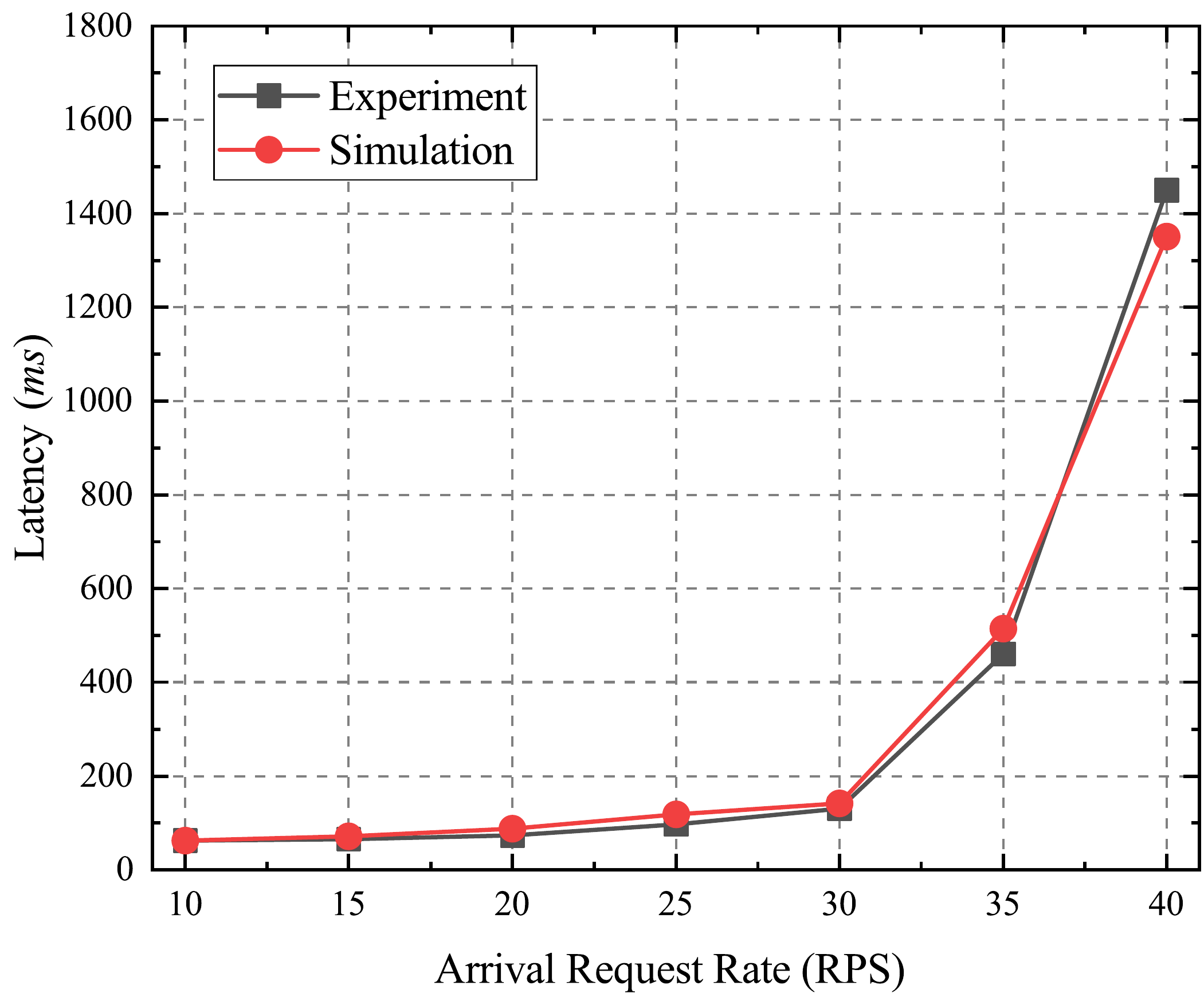}
		\caption{$N=1$}
	\end{subfigure}
	~ 
	\begin{subfigure}[b]{0.32\textwidth}
		\includegraphics[width=\textwidth]{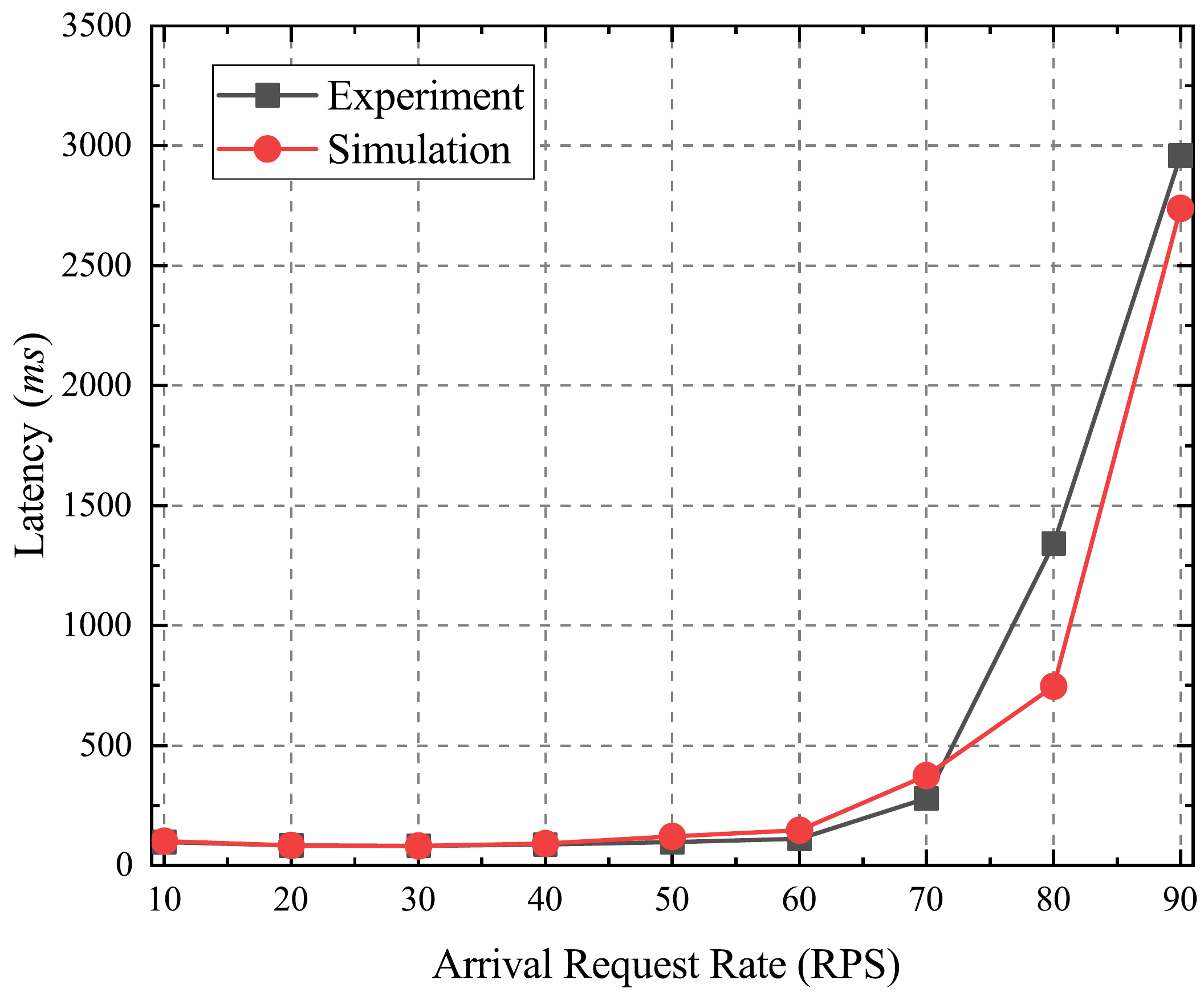}
		\caption{$N=2$}
	\end{subfigure}
	~ 
	\begin{subfigure}[b]{0.32\textwidth}
		\includegraphics[width=\textwidth]{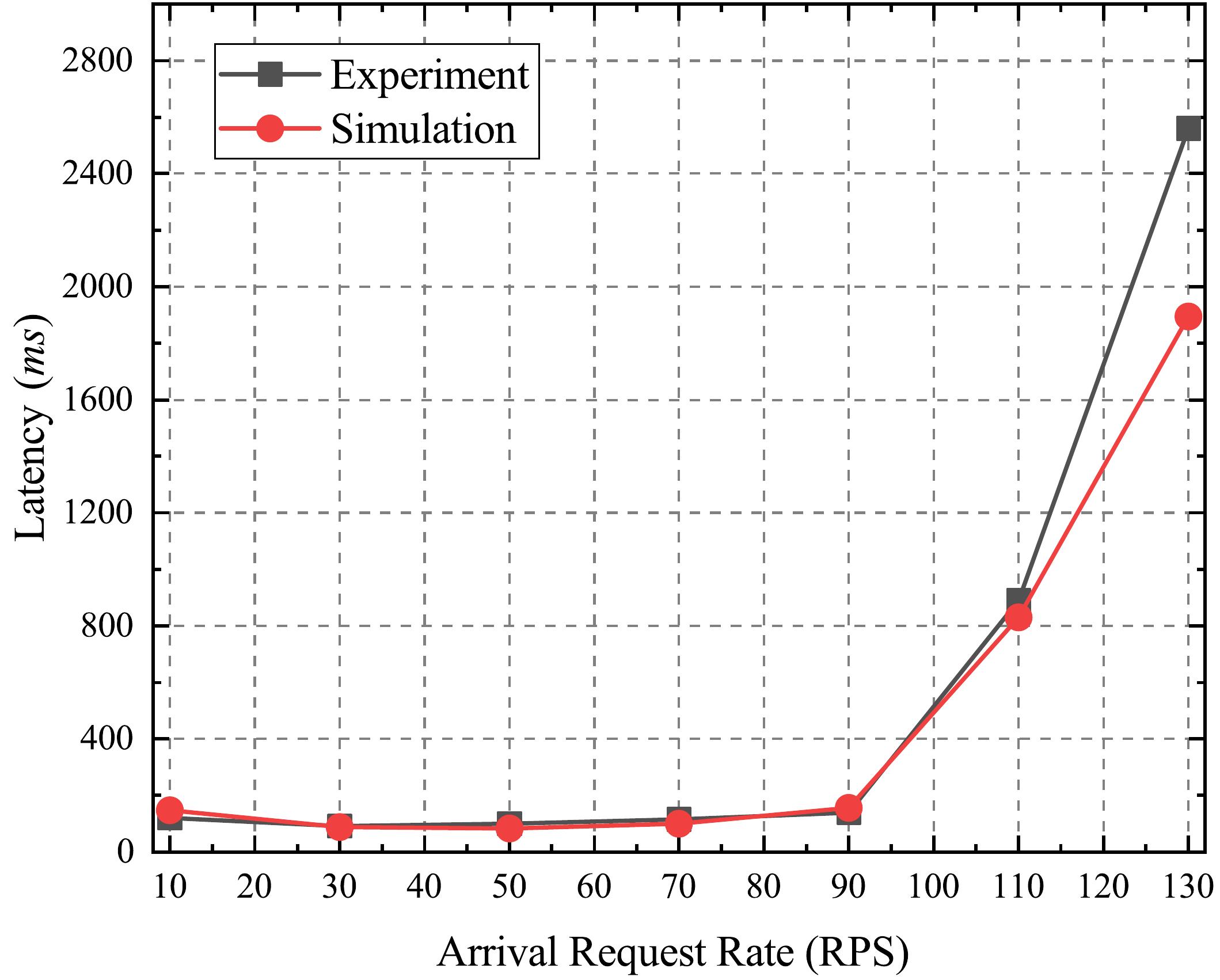}
		\caption{$N=3$}
	\end{subfigure}
	\caption{System Latency with Different $N$.}
\end{figure*}

\begin{figure*}
	\centering
	\begin{subfigure}[b]{0.32\textwidth}
		\includegraphics[width=\textwidth]{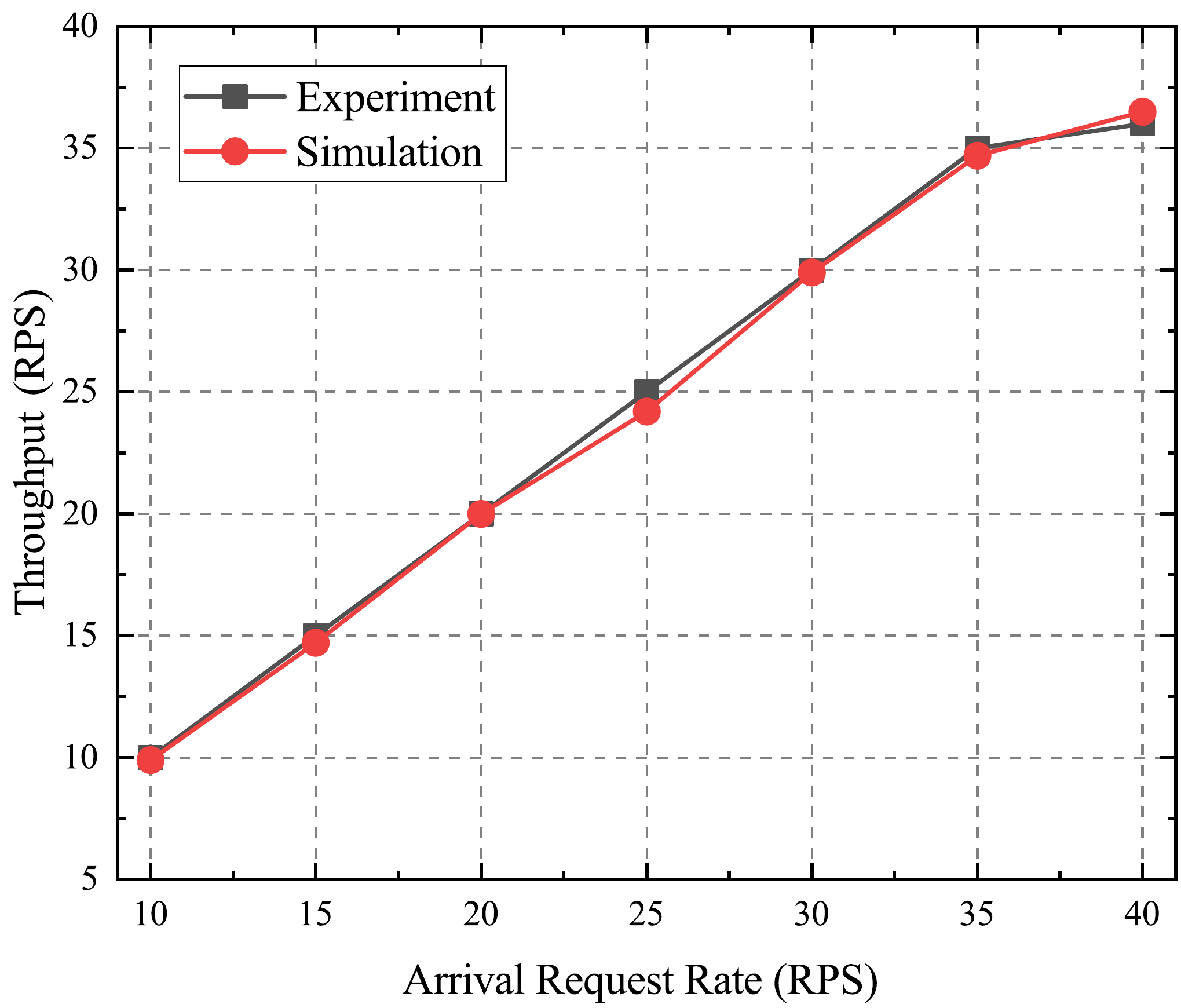}
		\caption{$N=1$}
	\end{subfigure}
	~ 
	\begin{subfigure}[b]{0.32\textwidth}
		\includegraphics[width=\textwidth]{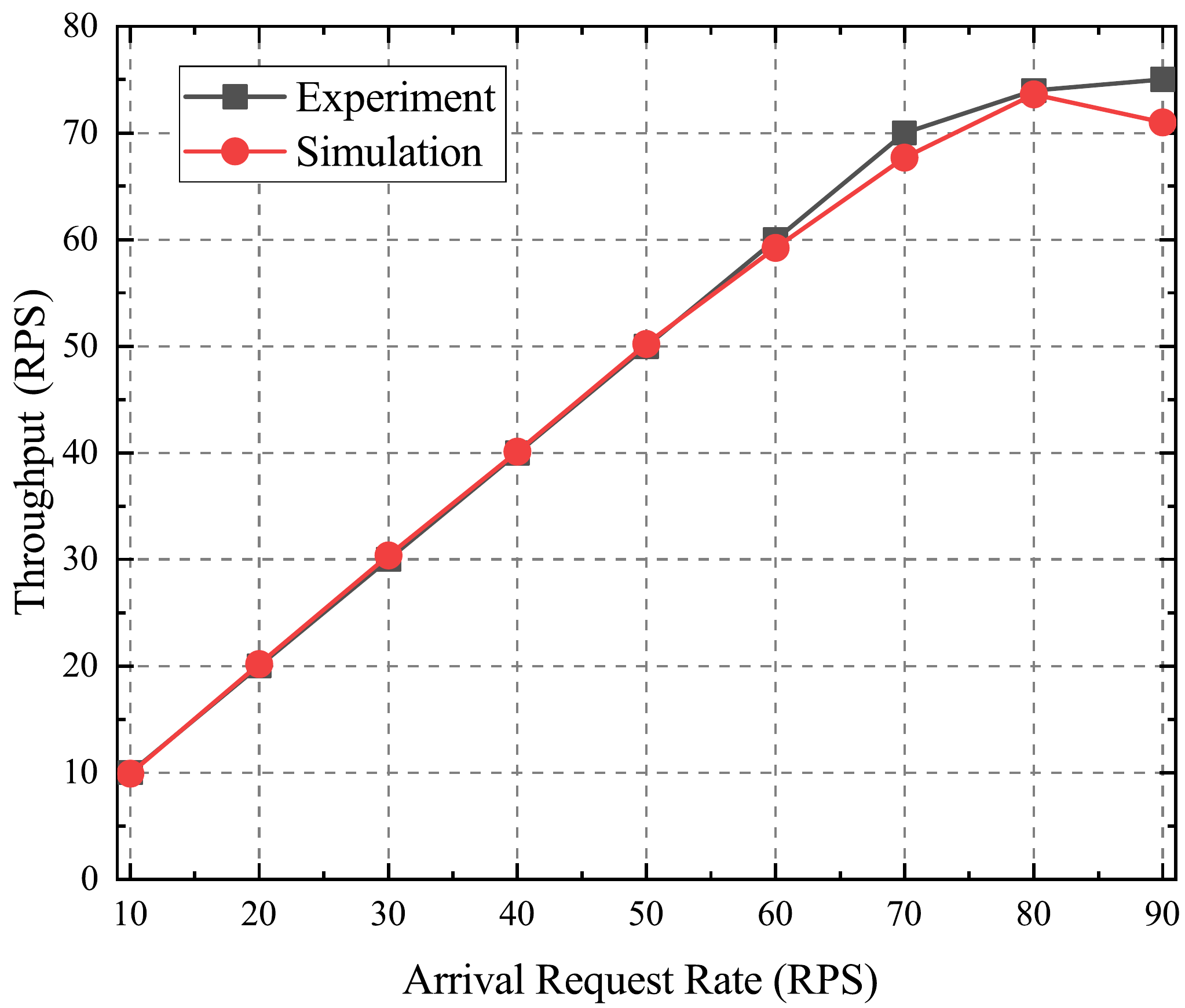}
		\caption{$N=2$}
	\end{subfigure}
	~ 
	\begin{subfigure}[b]{0.32\textwidth}
		\includegraphics[width=\textwidth]{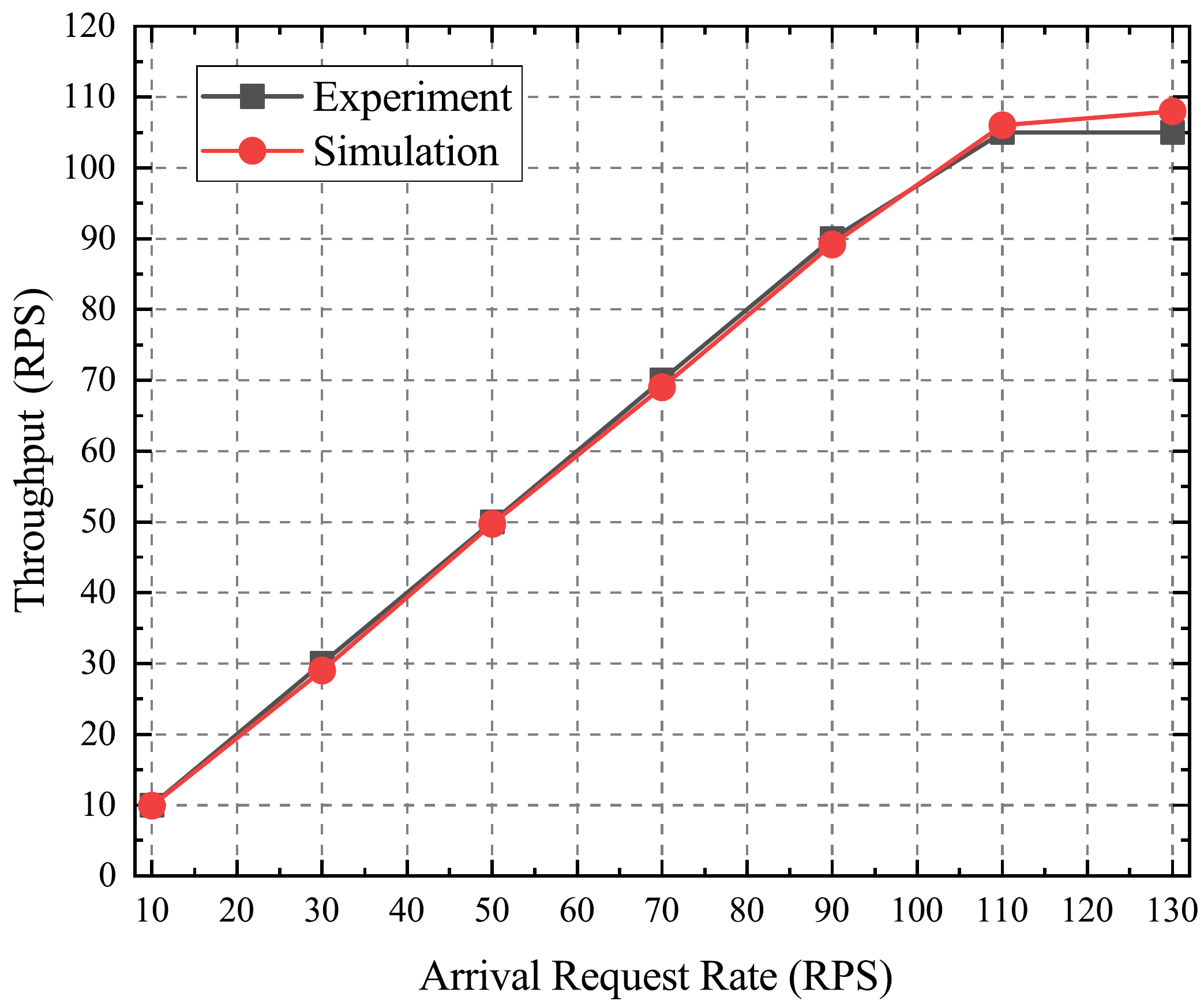}
		\caption{$N=3$}
	\end{subfigure}
	\caption{System Throughput with Different $N$.}
\end{figure*}
\subsection{Network Parameter Determination}
As we know, the most time-consuming operation in the ordering phase is signing and delivering the blocks \cite{b18}. It is feasible to reduce the block generation rate by adjusting the configuration of the ordering service, which can guarantee the ordering phase is not the performance bottleneck. Moreover, the experimental results in \cite{b24} indicate that at a high request arrival rate, the performance bottleneck of the Hyperledger Fabric framework exists in the endorsement phase or the committing phase (We will prove this in the next section). Moreover, the number of transactions contained in the block (or block size) has a great influence on the system performance. Under this premise, we can discuss the impact of the different parameters of ordering service, in order to find the best configuration parameters to optimize the system performance (i.e., make the system achieve the max throughput). The overall throughput is selected as the performance indicator rather than the overall latency and the reasons are listed as follows, i.e.,
\begin{itemize}
	\item Many related studies indicate that the throughput is more important than the latency in a Hyperledger-based system\cite{b18,b24}.
	\item With the increase of the requests arrival rate, the system latency can grow infinitely while the throughput can reach a saturation point.
	\item The overall latency of a Hyperledger-based system is greatly affected by the configuration parameters of the network and the request arrival rate. It is obvious that the overall latency obtained from experiments can’t represent the real processing time because it contains uncertain waiting time due to those different parameters. However, considering the throughput can ignore the impact of the request arrival rate because the throughput can reach a saturation point when the request arrival rate is big enough.
\end{itemize}

The proposed analytic model based on GSPN indicates that the blockchain system is composed of multiple successive M/M/1 networks. Thus, the system throughput is equal to the lowest throughput of those phases. Obviously, different $N$ in model can greatly affect the arrival rate of the committing phase, which further affects the system performance. Considering the ordering service configuration mentioned earlier, the number of transactions within a single block is determined by specific packaging strategies. Assume that $\lambda$ denotes the arrival rate of requests, $t$ denotes the BatchTimeout, $n$  denotes the MaxMessageCount, $\mu_e$  denotes the service rate of endorsement, $\mu_c$ denotes the service rate of committing and $\mu_c$ is determined by $N$ with function $f(N)$. Our work is to find appropriate $n$ and $t$ to maximize the system throughput, i.e.,
\begin{equation}
\Theta_{max} =\max\min(N*f(N),\mu_e),
\end{equation}
where $N$ is determined by $n$ and $t$ with function $g(n,t)$, i.e.,
\begin{equation}
g(n,t)=
\begin{cases}
	n, & \lambda \ge \frac{n}{t},\\
	\lfloor t*\lambda \rfloor + 1, & 0 < \lambda < \frac{n}{t}.
\end{cases}\label{n}
\end{equation}

\begin{figure*}
	\centering
	\begin{subfigure}[b]{0.32\textwidth}
		\includegraphics[width=\textwidth]{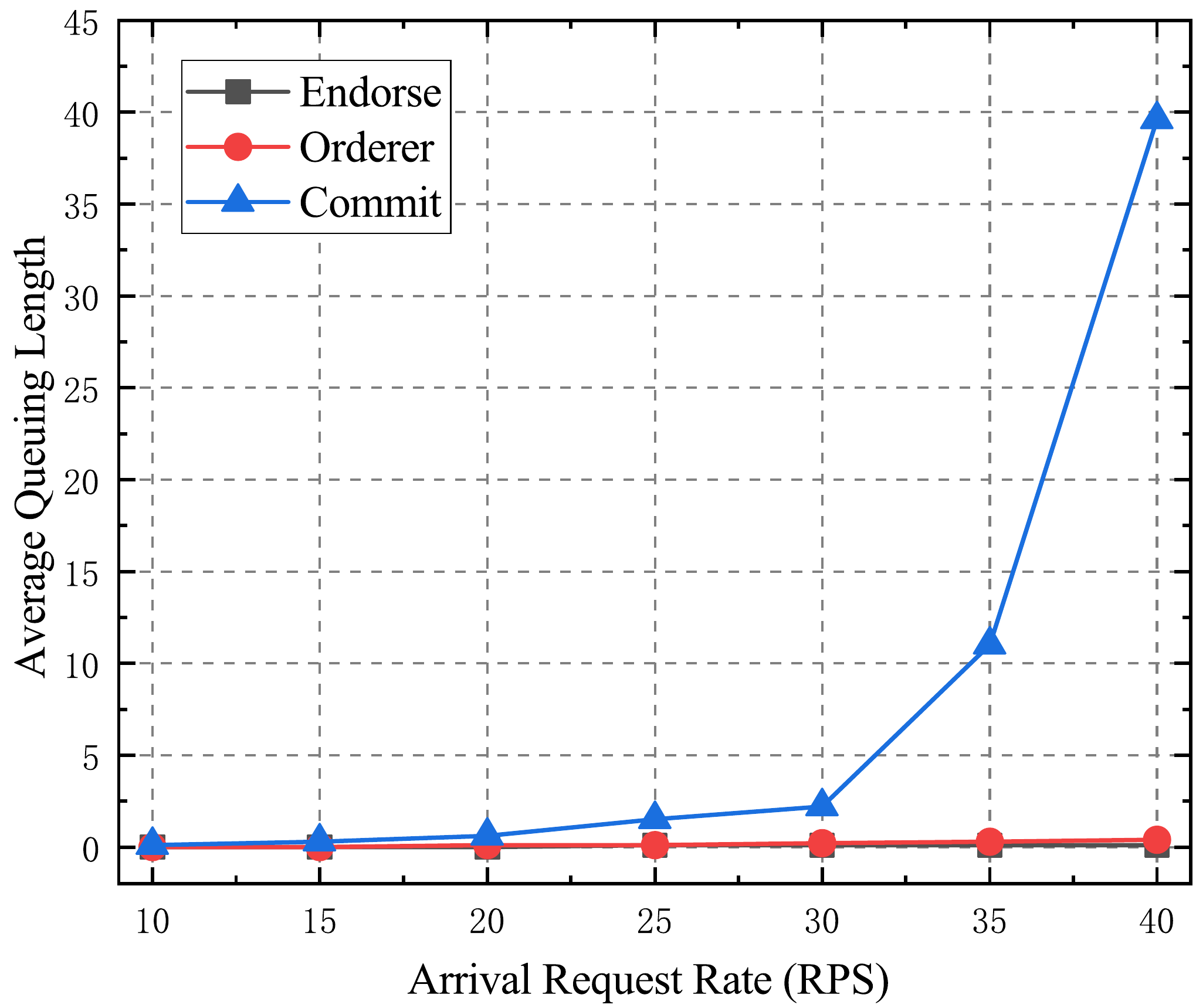}
		\caption{$N=1$}
	\end{subfigure}
	~ 
	\begin{subfigure}[b]{0.32\textwidth}
		\includegraphics[width=\textwidth]{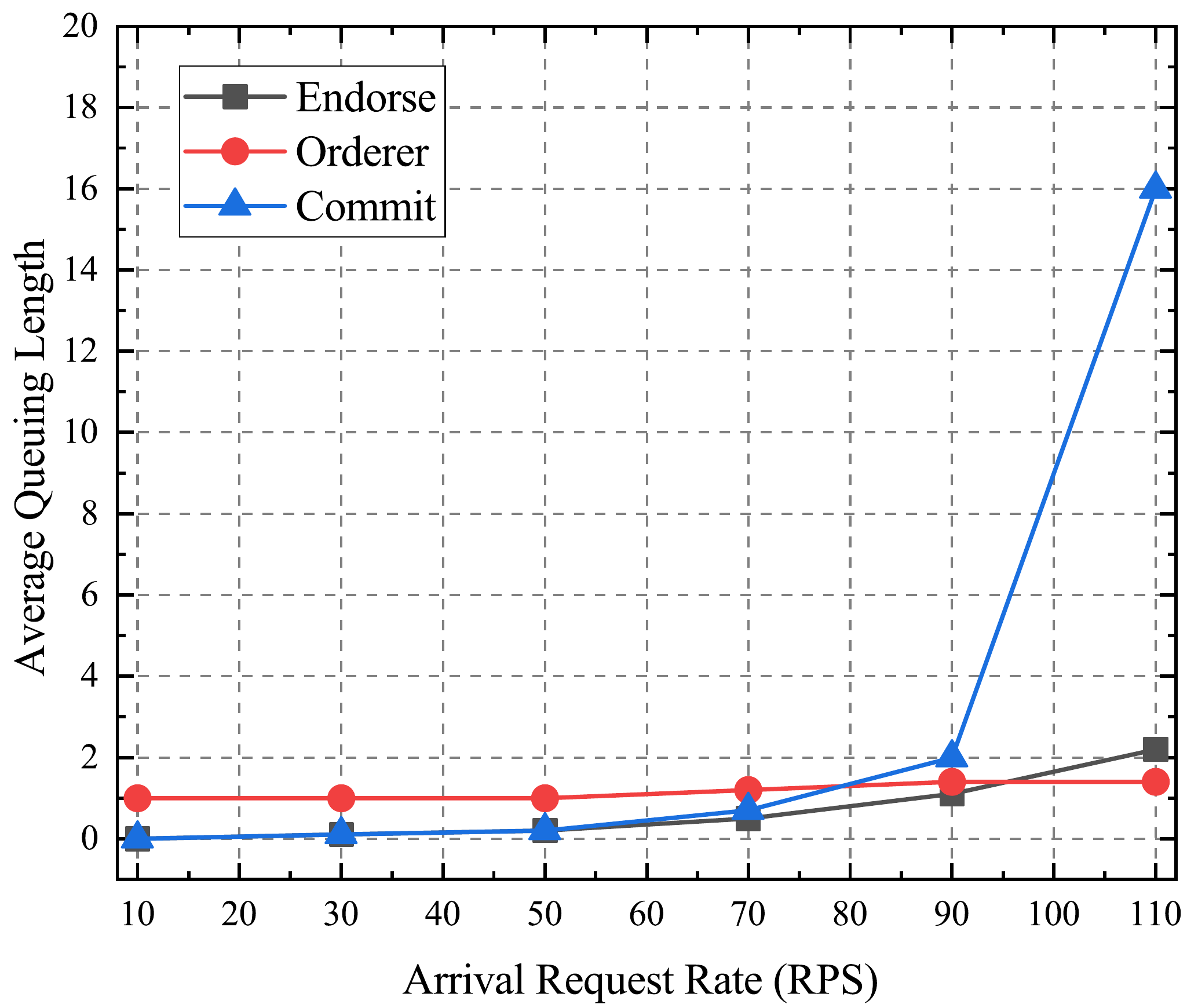}
		\caption{$N=3$}
	\end{subfigure}
	~ 
	\begin{subfigure}[b]{0.32\textwidth}
		\includegraphics[width=\textwidth]{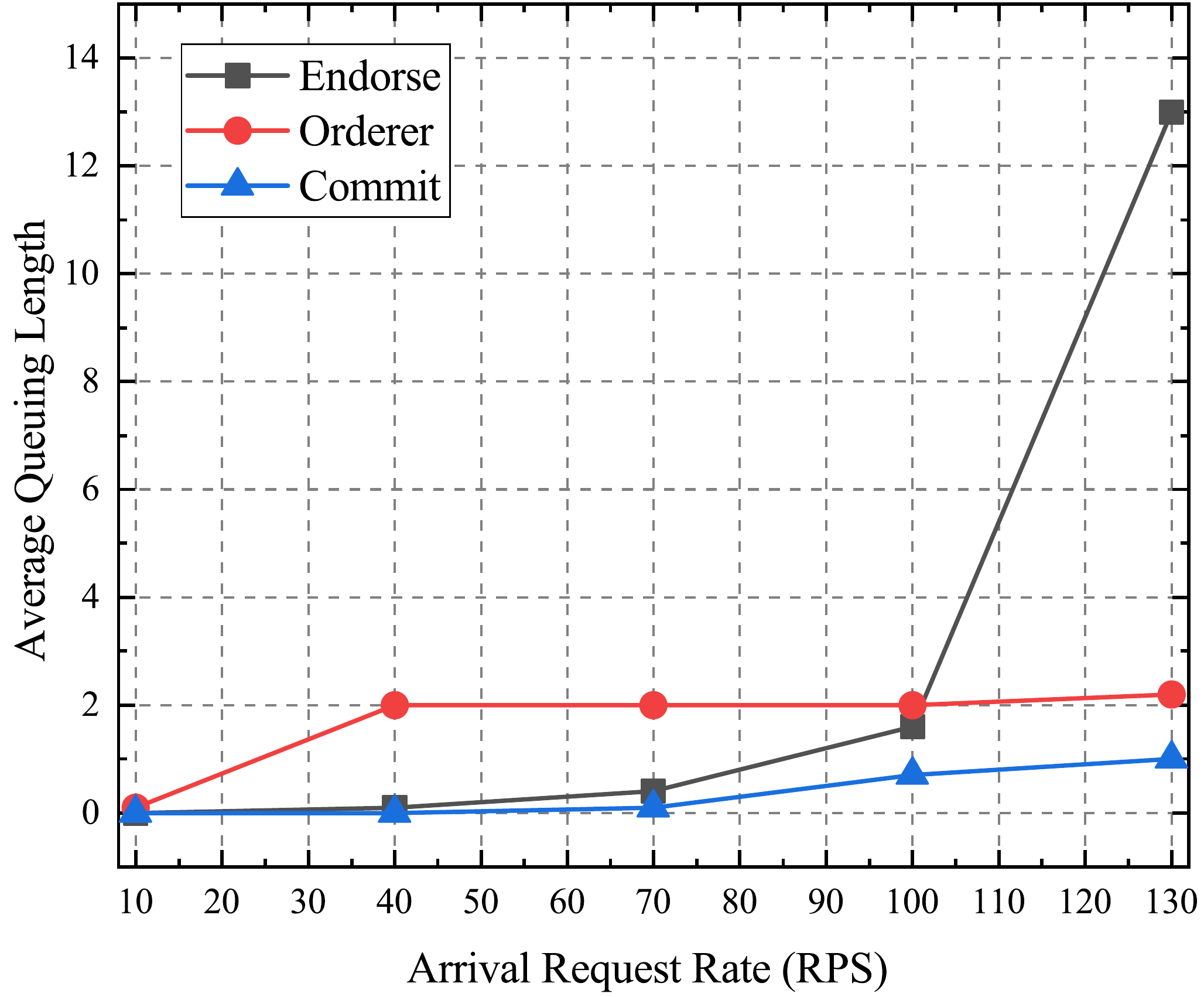}
		\caption{$N=5$}
	\end{subfigure}
    \caption{Queuing Length with Different $N$.}
\end{figure*}

\begin{figure*}
	\centering
	\begin{subfigure}[b]{0.32\textwidth}
		\includegraphics[width=\textwidth]{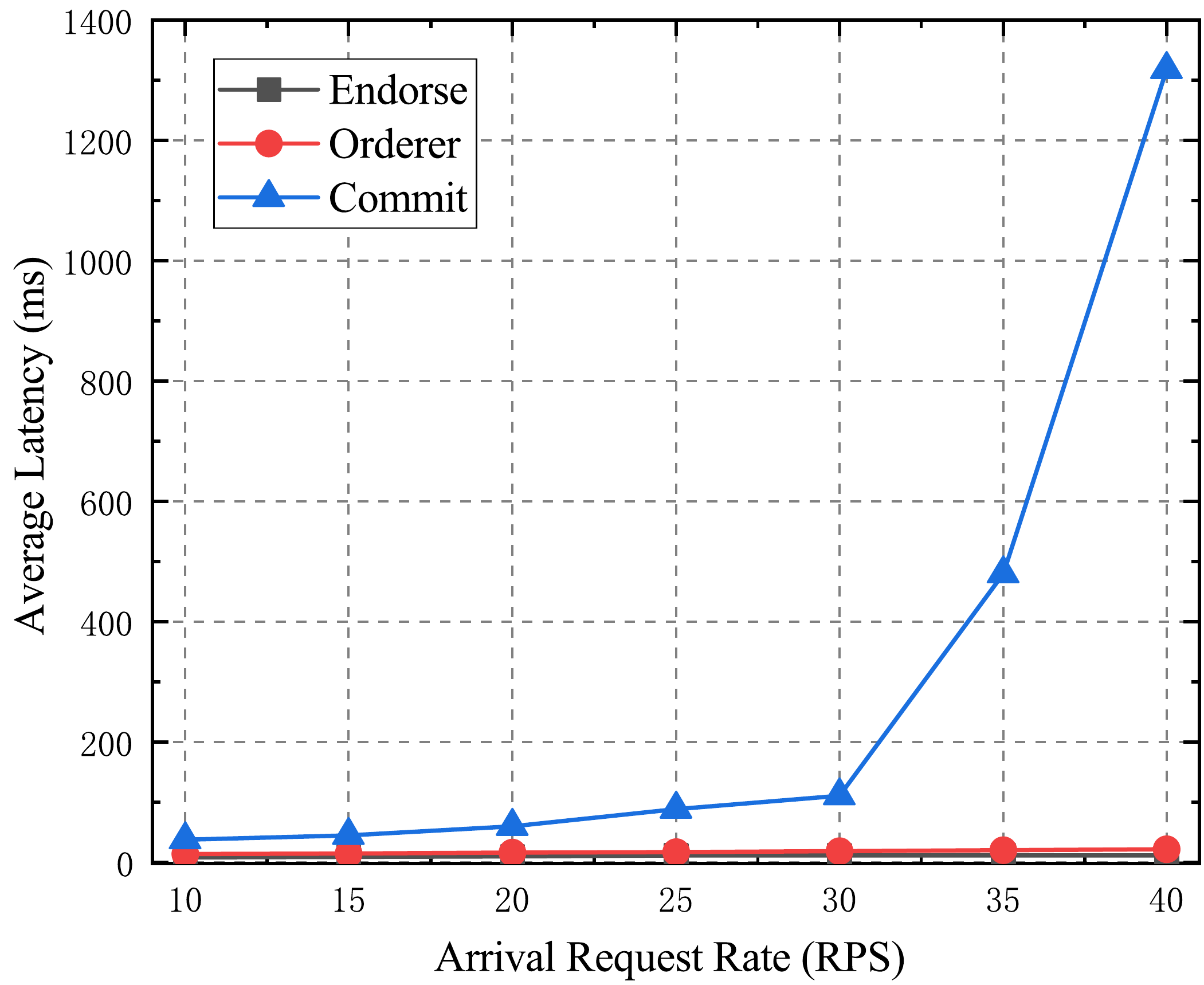}
		\caption{$N=1$}
	\end{subfigure}
	\begin{subfigure}[b]{0.32\textwidth}
		\includegraphics[width=\textwidth]{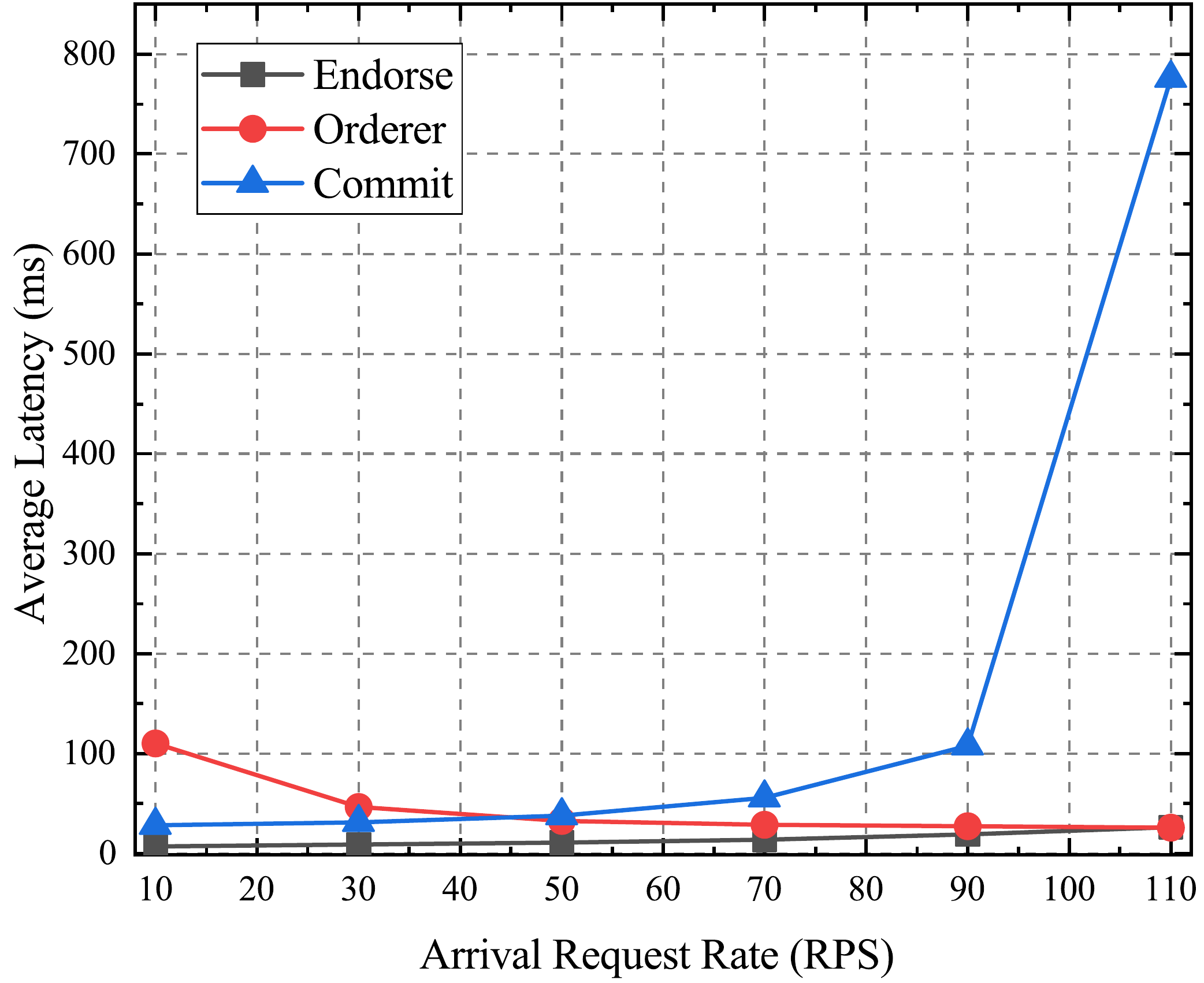}
		\caption{$N=3$}
	\end{subfigure}
	\begin{subfigure}[b]{0.32\textwidth}
		\includegraphics[width=\textwidth]{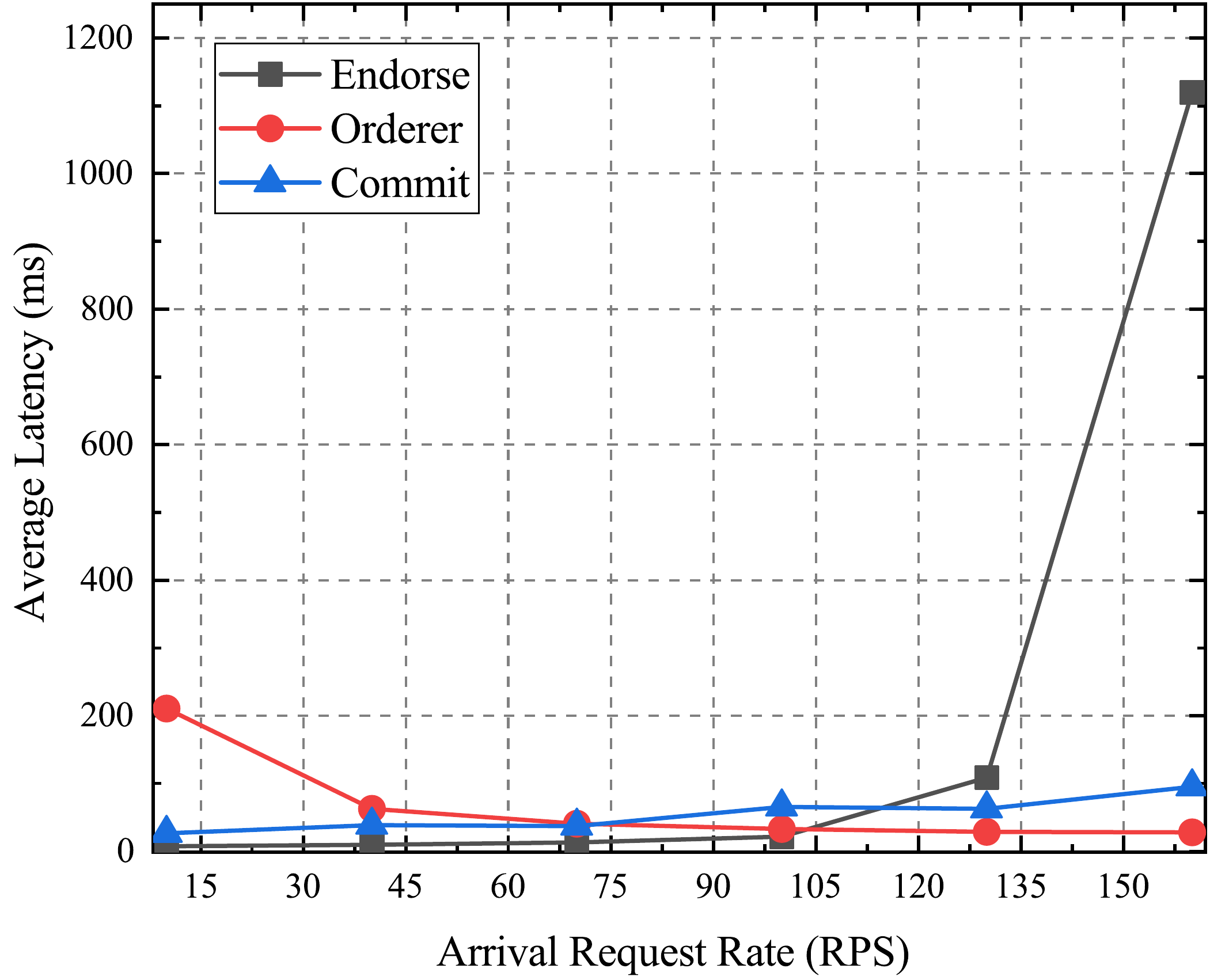}
		\caption{$N=5$}
	\end{subfigure}
    \caption{Average Latency of Each Phase with Different $N$.}
\end{figure*}

It is clear that we can adjust $N$ to make sure the throughput reaches the maximum value $\mu_e$, so there is $\lambda \ge \mu_e$. Considering that $\lambda$ is an uncertain value in the practical use, it is inappropriate to determine the configuration parameters depending on it, so it is better to assign it to a constant value. It is obvious that the obtained $n$ and $t$ under the premise $\lambda=\mu_e$ can still make sure the system throughput reach $\mu_e$ when $\lambda>\mu_e$, thus \eqref{n} can be simplified by taking $\lambda=\mu_e$. Moreover, $n$ and $t$ can be considered respectively, which ignores the impact of $\lambda$ and guarantees $\mu_e$ is always the smallest one in the formula, so that $t$ can not be the bottleneck (e.g. when $t$ and $\lambda$ are small, the number of transactions arrived in $t$ may be less than $n$, thus $N$ is limited). Therefore,

\begin{equation}
\begin{split}
\Theta_{max} &=\max\min(\mu_e, n*f(n),\\
&(\lfloor t*\mu_e \rfloor + 1)*f(\lfloor t*\mu_e \rfloor + 1)).\label{max}
\end{split}
\end{equation}
Then, after determining $\mu_e$ and $f(N)$, the appropriate configuration parameters can be obtained to achieve the maximum throughput of the system.

\section{Performance Evaluation \& Optimization}
\subsection{System Configuration}
To validate the proposed model, the system mentioned in Section III is set up on a aliyun cloud server with 4 Intel(R) Xeon(R) CPU E5-2680 v3 @ 2.50GHz processors and 8GB RAM, running Ubuntu 14.04 (64 bit). Various experiments are conducted on this running system by sending HTTP requests.
To simulate the GSPN-based model needs to determine the firing time of all the transitions at first. For the HTTP phase and endorsement phase, the latency can be calculated by printing timestamp at the specific code point in HTTP server. For ordering phase and committing phase, the time information can be extracted from log files after adjusting the log level of relevant docker containers from INFO to DEBUG. Finally, for response phase, an extra empty interface is integrated in the HTTP server to obtain the end-to-end latency. The average latency of all phases are summarized in Table I.
\begin{table}[ht]
	\centering
	\setlength{\tabcolsep}{4.2mm}%
	\caption{Model Parameters.}
	\label{Model Parameters}
	\begin{tabular}{cccccc}
		\hline
		\multicolumn{1}{|c|}{Transision}   & \multicolumn{1}{c|}{T$_h$}   & \multicolumn{1}{c|}{T$_e$}   & \multicolumn{1}{c|}{T$_o$}  & \multicolumn{1}{c|}{T$_c$}  & \multicolumn{1}{c|}{T$_n$}   \\ \hline
		\multicolumn{1}{|c|}{Latency (ms)} & \multicolumn{1}{c|}{2}   & \multicolumn{1}{c|}{7}   & \multicolumn{1}{c|}{12} & \multicolumn{1}{c|}{27} & \multicolumn{1}{c|}{10}  \\ \hline
		\multicolumn{1}{|c|}{Rate}         & \multicolumn{1}{c|}{500} & \multicolumn{1}{c|}{143} & \multicolumn{1}{c|}{83} & \multicolumn{1}{c|}{37} & \multicolumn{1}{c|}{100} \\ \hline
		\multicolumn{1}{l}{}               & \multicolumn{1}{l}{}     & \multicolumn{1}{l}{}     & \multicolumn{1}{l}{}    & \multicolumn{1}{l}{}    & \multicolumn{1}{l}{}    
	\end{tabular}
\end{table}

\subsection{Model Validation}
Based on the running system, Locust\footnote{https://www.locust.io/} is used to evaluate the system overall latency and throughput of HTTP interfaces over increasing request arrival rates and different $N$. The throughput is evaluated by Requests Per Second (RPS). Moreover, a software tool embedded in Matlab named pntool\footnote{http://www.pntool.ac.tuiasi.ro/} is used to simulate the GSPN-based model under the same conditions. This tool can obtain the performance metrics (e.g., latency, throughput, queuing length) for all the transitions and places of the model. By analyzing the simulation results, the overall latency and throughput can be determined. Fig. 5 and Fig. 6 compare the experimental results with the simulation results of the system. It can be observed that the experimental results are comparable to the simulation results and the proposed model can well describe the actual system. Furthermore, the overall latency increases rapidly when throughput reaches the saturation point and the maximum throughput is greatly improved due to the increase of $N$, which indicates that the different configuration parameters ($N$) has a great influence on the system performance.

\subsection{System Bottleneck Analysis}
System bottleneck analysis is significant for performance improvement. For a Hyperledger-based system, once the bottleneck (i.e., endorsement phase, ordering phase and committing phase) is identified, the optimization goal is determined. It is very hard to obtain the performance metrics of each processing phase through experiments as the official SDK does not provide corresponding APIs for developers. Thus, we analyze the system bottleneck based on the proposed GSPN model which has been validated. Because each place in this model represent a specific processing phase, the performance metrics for each processing phase such as the average latency can be obtained by simulating the model. Consider the average latency and queuing length of each phase (i.e., places like $P_{wait\_h}$) as the metrics of system busyness, the longer queuing length and latency lead to the worse performance. Fig. 7 shows the average queuing length of different phases with different values of $N$, i.e., 1, 3 and 5. When $N$ is small, the queuing length of committing phase increases significantly while the other two curves remain stable. However, when $N=5$, the queuing length of endorsement phase increases rapidly. This is because the increase of $N$ is equivalent to decreasing the arrival rate of the committing phase. When $N$ is large enough, the service rate of the committing phase is more than the arrival rate of requests while the endorsement phase is less than it.  Fig. 8 shows the average latency of the three major processing phases with different $N$, and the results match the previous analysis. Therefore, it can be found that the committing phase is the system bottleneck in case of small $N$ while the endorsement phase is instead in case of large $N$. Thus, it is an effective approach to improve the system performance by determining the value of $N$ in the practical use of a Hyperledger-based system.
\subsection{Performance Optimization}
The performance metrics in Fig. 6 point out that the value of $N$ has a great influence on the maximum throughput of the Hyperledger-based system. In Section IV, a mathematical method is proposed to determine the configuration parameters of ordering service, in order to achieve the maximum throughput. Considering the analysis results of the system bottleneck in the previous subsection, the proposed method is actually an approach to improve the performance  of the committing phase. According to \eqref{max}, when the function $f(N)$ is determined,  the appropriate $n$ and $t$ can be obtained. At this point, the system bottleneck is the endorsement phase.

We leverage the approach of curve fitting to approximate $f(N)$. The experimental data and fitting result are shown in Fig. 9. It is obvious that the latency of committing phase is linear with $N$ within a certain range and the function is
\begin{equation}
h(N)=1000/f(N)=25.06+1.57*N.
\end{equation}
Therefore, the maximum throughput of the system is
\begin{equation}
\Theta_{max}=\max\min(\mu_e, \frac{1000}{\frac{25.06}{n}+1.57},
\frac{1000}{\frac{25.06}{\lfloor t*\mu_e \rfloor + 1}+1.57}).
\end{equation}
To solve (12), we have
\begin{equation}
\Theta_{max}=\mu_e.
\end{equation}
\begin{equation}
n \ge \lceil \frac{25.06*\mu_e}{1000-1.57*\mu_e} \rceil, t \ge \frac{26.63*\mu_e - 1000}{\mu_e * (1000-1.57*\mu_e)}.
\end{equation}
According to the results in table I, let $\mu_e=143$. Thus, the theoretical results show that the system can reach a maximum throughput of 143 RPS when $n \ge 5$ and $t \ge 0.026$ (s). Note that $t$ is determined under the premise of $\lambda=\mu_e$, a larger $\lambda$ can ignore the impact of $t$. However, restricting the range of $t$ can prevent $t$ from being the bottleneck. Generally, it is not suggested to assign $n$ and $t$ with a large value in the practical environment because the large value can lead to poor performance with a low arrival rate $\lambda$.

To validate the theoretical results above, a series of experiments have been conducted on the running system. For each $N$, we gradually increased the arrival rate in Locust until the throughput reaches the saturation point. Fig. 10 depicts the relationship between $N$ and the maximum system throughput and compares the simulation and experimental results. Our goal is to determine the maximum throughput and the corresponding $N$. The experimental results show that when $N$ is larger than 5, the maximum throughput becomes floor at around 145 RPS. Thus, continuing to increase $N$ when $N \ge 5$ can not significantly improve the throughput, which is in line with the theoretical results (i.e., 143 RPS when $n \ge 5$ and $t \ge 0.026$ (s)). Therefore, our proposed approach can determine the appropriate configuration of ordering service. As for other Hyperledger-based systems with distinct underlying networks, the proposed analytic method can be also applied to achieve the higher system throughput. 

The steps to determine the configuration parameters are described as follows, i.e.,:
\begin{itemize}
	\item Step 1: Obtain the service rate of endorsement phase ($\mu_e$) and committing phase ($\mu_c$) by conducting experiments on the system.
	\item Step 2: Determine which processing phase of the transaction flow is the system bottleneck (The endorsement phase in our case).
	\item Step 3: Calculate $t$ (BatchTimeout) and $n$ (MaxMessageCount) by the proposed formula.
	\item Step 4: Confirm the conclusion through experiments.
\end{itemize}

Actually, this analysis method can be applied to a more complex system with more organizations and channels. The only difference is how to get the $\mu_e$  and $\mu_c$ because the number of channels and peers can affect the service rate. Thus, even though the system in the paper only contains two organizations and one channel, the proposed analysis method is scalable and convincing.

\begin{figure}[htbp]
	\centerline{\includegraphics[width=\linewidth]{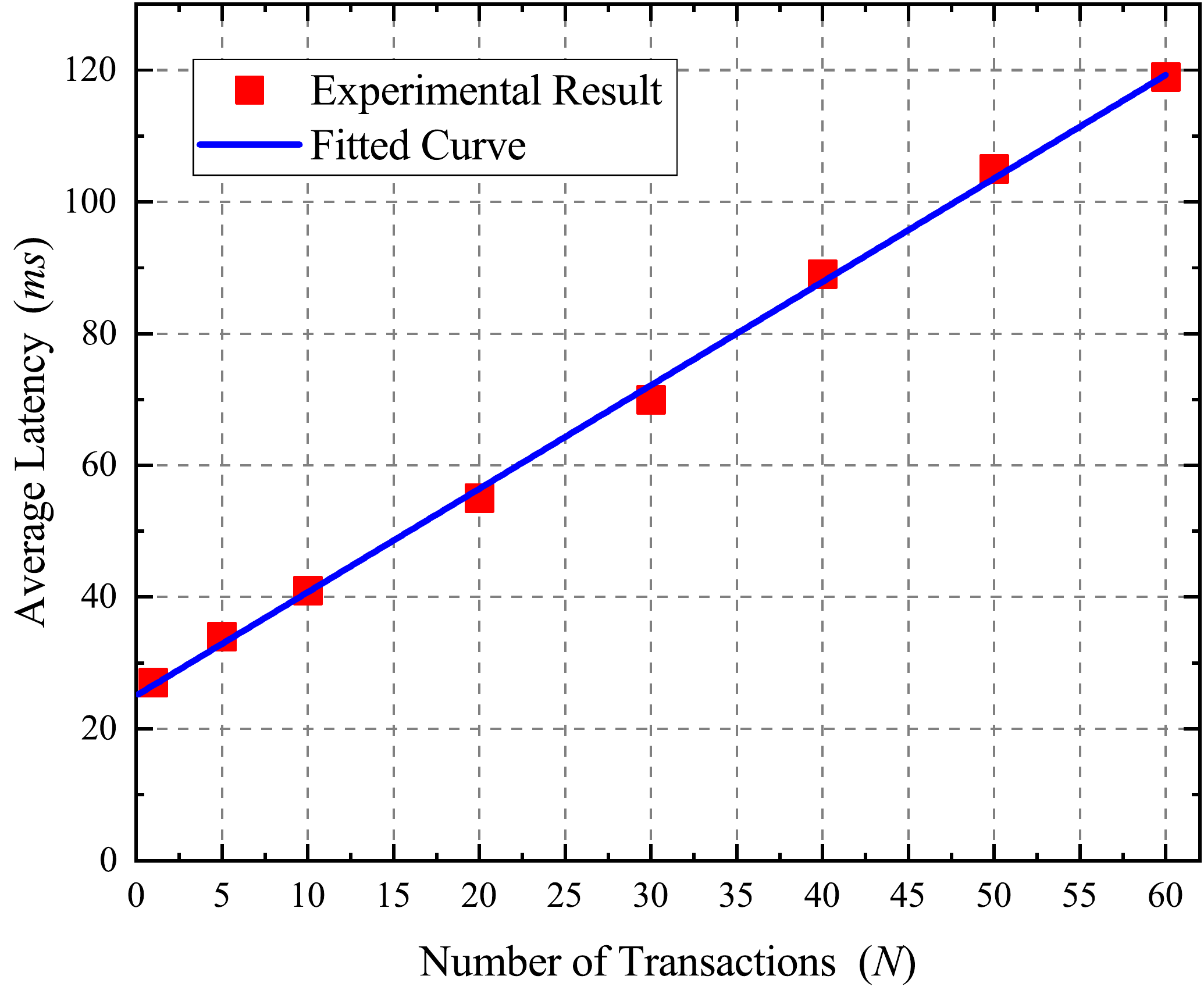}}
	\caption{Latency of Committing Phase with Different $N$.}
	\label{fig9}
\end{figure}

\begin{figure}[htbp]
	\centerline{\includegraphics[width=\linewidth]{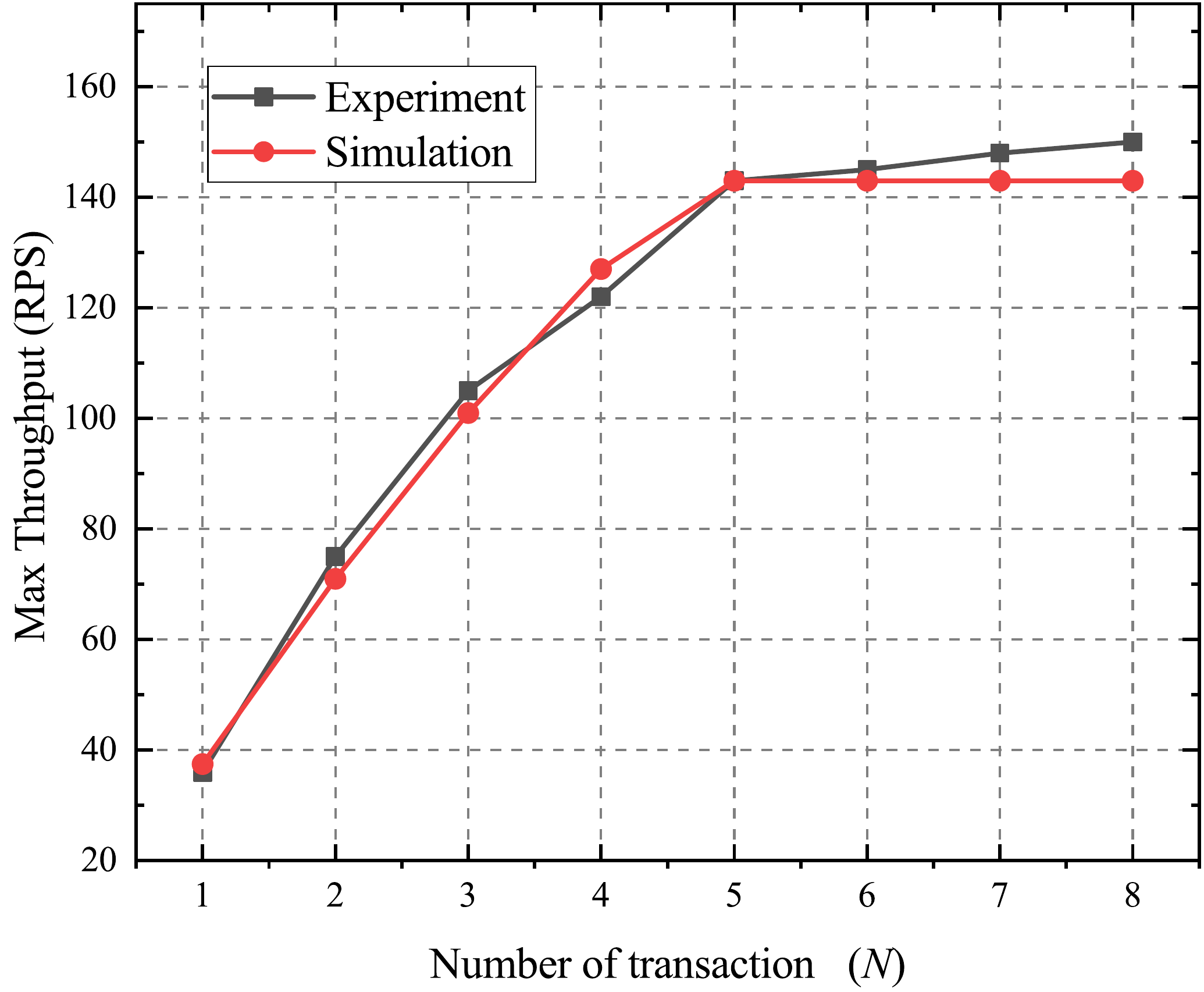}}
	\caption{Maximum throughput with Different $N$.}
	\label{fig10}
\end{figure}

\section{Conclusion}
In this paper, we have proposed a GSPN model for a blockchain system based on Hyperledger Fabric v1.2. This model depicts the transaction flow of Hyperledger Fabric in details and is aiming to evaluate the system performances. Extensive experiments have been conducted on a real-time system to validate our model. The results show that the number of transactions in a block significantly affect the system performance. Based on this model, we have analyzed the performance bottleneck with different configurations of the ordering service. With an increasing number of transactions in a block, the system bottleneck changes from committing phase to endorsement phase. In addition, we have presented a configuration selection approach to determine the configuration parameters of ordering service, in order to achieve the maximum throughput. Our simulation results have shown that when the number of transactions contained in a block exceeds 5 and the waiting time exceeds 0.26 seconds, the throughput can reach 143 RPS, which is in line with the real system. Furthermore, the conclusions of this paper are instructive for the future work. Our next plan is to further improve the overall system performance by optimizing the endorsement phase. For example, adding more endorsers in each organization to load balance the endorsement is an effective approach.

\section*{Acknowledgment}
This work is supported by Chinese National Natural Science Foundation (61671089) and China Unicom Network Technology Research Institute.


\begin{thebibliography}{00}
\bibitem{b1} Nakamoto, ``Bitcoin: A peer-to-peer electronic cash system,'' White Paper, 2008.

\bibitem{b2} Z. Yang, K. Yang, L. Lei, K. Zheng and V. C. M. Leung, ``Blockchain-Based Decentralized Trust Management in Vehicular Networks,'' in IEEE Internet of Things Journal, vol. 6, no. 2, pp. 1495-1505, April 2019.

\bibitem{b3} A. Azaria, A. Ekblaw, T. Vieira and A. Lippman, ``MedRec: Using Blockchain for Medical Data Access and Permission Management,'' 2016 2nd International Conference on Open and BigData (OBD), Vienna, 2016, pp. 25-30.

\bibitem{b4} A. Stanciu, ``Blockchain Based Distributed Control System for Edge Computing,'' 2017 21st International Conference on Control Systems and Computer Science (CSCS), Bucharest, 2017, pp. 667-671.

\bibitem{b5} X. Xiong, K. Zheng, R. Xu, W. Xiang and P. Chatzimisios, ``Low power wide area machine-to-machine networks: key techniques and prototype," in IEEE Communications Magazine, vol. 53, no. 9, pp. 64-71, September 2015.

\bibitem{b6} F. Liu, K. Zheng, W. Xiang and H. Zhao, ``Design and Performance Analysis of An Energy-Efficient Uplink Carrier Aggregation Scheme," in IEEE Journal on Selected Areas in Communications, vol. 32, no. 2, pp. 197-207, February 2014.

\bibitem{b7} Sachin S. Shetty; Charles A. Kamhoua; Laurent L. Njilla, ``Permissioned and Permissionless Blockchains,'' in Blockchain for Distributed Systems Security, IEEE, 2019, pp.193-204.

\bibitem{b8} S. Wang, Y. Yuan, X. Wang, J. Li, R. Qin and F. Wang, ``An Overview of Smart Contract: Architecture, Applications, and Future Trends,'' 2018 IEEE Intelligent Vehicles Symposium (IV), Changshu, 2018, pp. 108-113.

\bibitem{b9} M. Wohrer and U. Zdun, ``Smart contracts: security patterns in the ethereum ecosystem and solidity,'' 2018 International Workshop on Blockchain Oriented Software Engineering (IWBOSE), Campobasso, 2018, pp. 2-8.

\bibitem{b10} Elli Androulaki, Artem Barger, Vita Bortnikov, et al,. 2018. ``Hyperledger fabric: a distributed operating system for permissioned blockchains.'' In Proceedings of the Thirteenth EuroSys Conference (EuroSys ’18). ACM, New York, NY, USA, Article 30, 15 pages.

\bibitem{b11} P. Yuan, X. Xiong, L. Lei and K. Zheng, ``Design and Implementation on Hyperledger-Based Emission Trading System,'' in IEEE Access, vol. 7, pp. 6109-6116, 2019.

\bibitem{b12} M. Raikwar, S. Mazumdar, S. Ruj, S. S. Gupta, A. Chattopadhyay, K. Lam, ``A blockchain framework for insurance processes'', Proc. 9th IFIP Int. Conf. New Technol. Mobility Secur. (NTMS), pp. 1-4, 2018. 

\bibitem{b13} Q. Liu, Q. Guan, X. Yang, H. Zhu, G. Green and S. Yin, ``Education-Industry Cooperative System Based on Blockchain,'' 2018 1st IEEE International Conference on Hot Information-Centric Networking (HotICN), Shenzhen, 2018, pp. 207-211.

\bibitem{b14} Q. Zheng, K. Zheng, H. Zhang and V. C. M. Leung, ``Delay-Optimal Virtualized Radio Resource Scheduling in Software-Defined Vehicular Networks via Stochastic Learning," in IEEE Transactions on Vehicular Technology, vol. 65, no. 10, pp. 7857-7867, Oct. 2016.

\bibitem{b15} A. Baliga, N. Solanki, S. Verekar, A. Pednekar, P. Kamat and S. Chatterjee, ``Performance Characterization of Hyperledger Fabric,'' 2018 Crypto Valley Conference on Blockchain Technology (CVCBT), Zug, 2018, pp. 65-74. 

\bibitem{b16} T. T. A. Dinh, J. Wang, G. Chen, R. Liu, B. C. Ooi, and K.-L. Tan, ``BLOCKBENCH: A Framework for Analyzing Private Blockchains,'' in ACM International Conference on Management of Data, ser. SIGMOD, 2017, pp. 1085–1100.

\bibitem{b17} Qassim Nasir, Ilham A. Qasse, Manar Abu Talib, and Ali Bou Nassif, ``Performance Analysis of Hyperledger Fabric Platforms,'' Security and Communication Networks, vol. 2018, Article ID 3976093, 14 pages, 2018. https://doi.org/10.1155/2018/3976093. 

\bibitem{b18} S. Pongnumkul, C. Siripanpornchana and S. Thajchayapong, ``Performance Analysis of Private Blockchain Platforms in Varying Workloads,'' 2017 26th International Conference on Computer Communication and Networks (ICCCN), Vancouver, BC, 2017, pp. 1-6.

\bibitem{b19} Y. Hao, Y. Li, X. Dong, L. Fang and P. Chen, ``Performance Analysis of Consensus Algorithm in Private Blockchain,'' 2018 IEEE Intelligent Vehicles Symposium (IV), Changshu, 2018, pp. 280-285. 

\bibitem{b20} Thakkar, Parth, Senthil Nathan, and Balaji Vishwanathan. ``Performance Benchmarking and Optimizing Hyperledger Fabric Blockchain Platform.'' arXiv preprint arXiv:1805.11390 (2018). 

\bibitem{b21} L. Lei, Y. Zhang, X. S. Shen, C. Lin and Z. Zhong, ``Performance Analysis of Device-to-Device Communications with Dynamic Interference Using Stochastic Petri Nets,'' in IEEE Transactions on Wireless Communications, vol. 12, no. 12, pp. 6121-6141, December 2013.

\bibitem{b22} K. Zheng, F. Liu, L. Lei, C. Lin and Y. Jiang, ``Stochastic Performance Analysis of a Wireless Finite-State Markov Channel," in IEEE Transactions on Wireless Communications, vol. 12, no. 2, pp. 782-793, February 2013.

\bibitem{b23} H. Wang, L. Lei and K. Zheng, ``Flow-level performance analysis of random wireless network using stochastic Petri Nets,'' 2016 23rd International Conference on Telecommunications (ICT), Thessaloniki, 2016, pp. 1-5. 

\bibitem{b24} K. Zheng, H. Meng, P. Chatzimisios, L. Lei and X. Shen, ``An SMDP-Based Resource Allocation in Vehicular Cloud Computing Systems," in IEEE Transactions on Industrial Electronics, vol. 62, no. 12, pp. 7920-7928, Dec. 2015.

\bibitem{b25} H. Sukhwani, J. M. Martínez, X. Chang, K. S. Trivedi and A. Rindos, ``Performance Modeling of PBFT Consensus Process for Permissioned Blockchain Network (Hyperledger Fabric),'' 2017 IEEE 36th Symposium on Reliable Distributed Systems (SRDS), Hong Kong, 2017, pp. 253-255.

\bibitem{b26} H. Sukhwani, N. Wang, K. S. Trivedi and A. Rindos, ``Performance Modeling of Hyperledger Fabric (Permissioned Blockchain Network),'' 2018 IEEE 17th International Symposium on Network Computing and Applications (NCA), Cambridge, MA, 2018, pp. 1-8.

\end{thebibliography}
\end{document}